# A Review of Detection, Evolution, and Data Reconstruction Strategies for False Data Injection Attacks in Power Cyber-Physical Systems

Xiaoyong Bo

(Electrical and Information Engineering College, Jilin Agricultural Science and Technology University, Jilin 132101, China)

**Abstract**: The integration of information and physical systems in modern power grids has heightened vulnerabilities to False Data Injection Attacks (FDIAs), threatening the secure operation of power cyber-physical systems (CPS). This paper reviews FDIA detection, evolution, and data reconstruction strategies, highlighting cross-domain coordination, multi-temporal evolution, and stealth characteristics. Challenges in existing detection methods, including poor interpretability and data imbalance, are discussed, alongside advanced state-aware and action-control data reconstruction techniques. Key issues, such as modeling FDIA evolution and distinguishing malicious data from regular faults, are identified. Future directions to enhance system resilience and detection accuracy are proposed, contributing to the secure operation of power CPS.

## 1 Introduction

As new-generation information technology deeply penetrates the power system [1-3], a multitude of electrical, sensing, and computational devices are interconnected via electrical and communication networks, transforming traditional systems centered on physical equipment into highly integrated Cyber-Physical Systems (CPS), i.e., power CPS. This integration of power CPS, underpinned by the physical network of primary electrical equipment for energy flow and the information network for secondary control and protection information flow, signifies a shift towards complex networks where power and information systems converge [4-7]. The architecture of the power system from the CPS perspective is depicted in Figure 1, encompassing generation, transmission, conversion, distribution, and utilization [8-12]. Measuring devices within the information domain transmit operational state data of the electrical grid to the dispatch data network, which then, through control centers, computing centers, and data centers, issues system control commands that modify the operational state of the physical electrical grid [13-17]. The evolution of power systems towards greater dependence on CPS is evident [18-21], but this integration also raises the susceptibility of the grid to cyber-attacks [22-25].

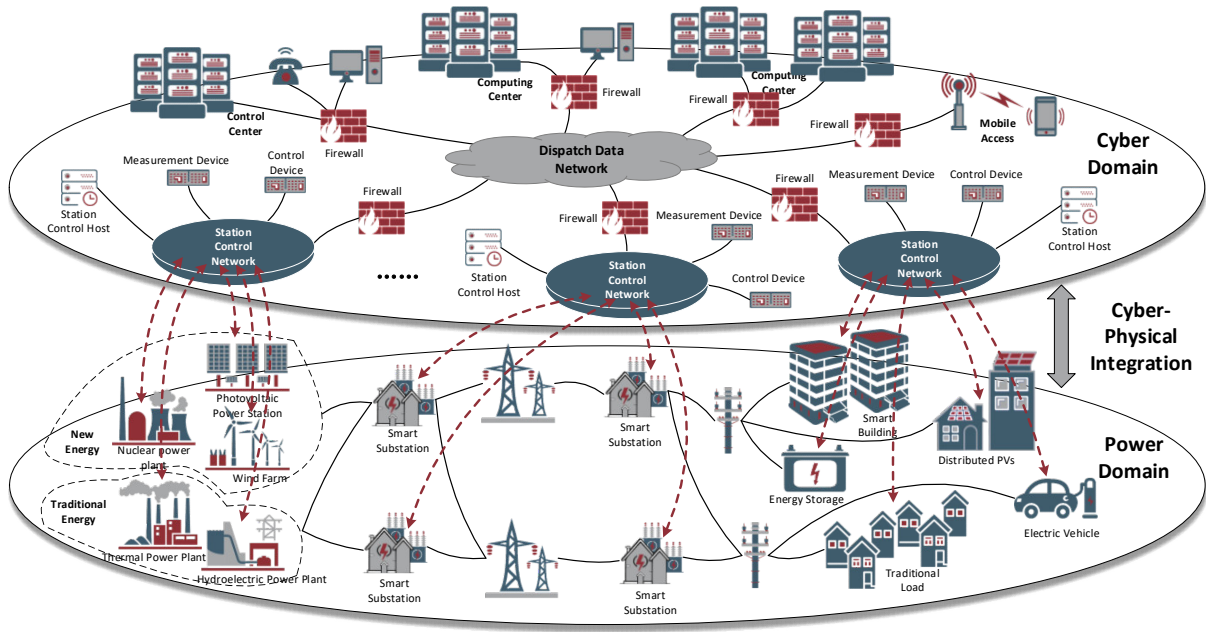

Fig. 1 Architecture of power system from the perspective of CPS

Recent years have seen an increase in cyber-attack incidents where hackers have infiltrated power grids, causing significant damage [26-28]. Examples include the Stuxnet virus attacking Iranian nuclear facilities in 2010, the BlackEnergy virus attack on the Ukrainian power grid in 2015, and the ARP cache poisoning at a U.S. wind farm in 2017, among others. These incidents are typical cases where the cyber-physical security of the power systems was compromised, leading to widespread outages.

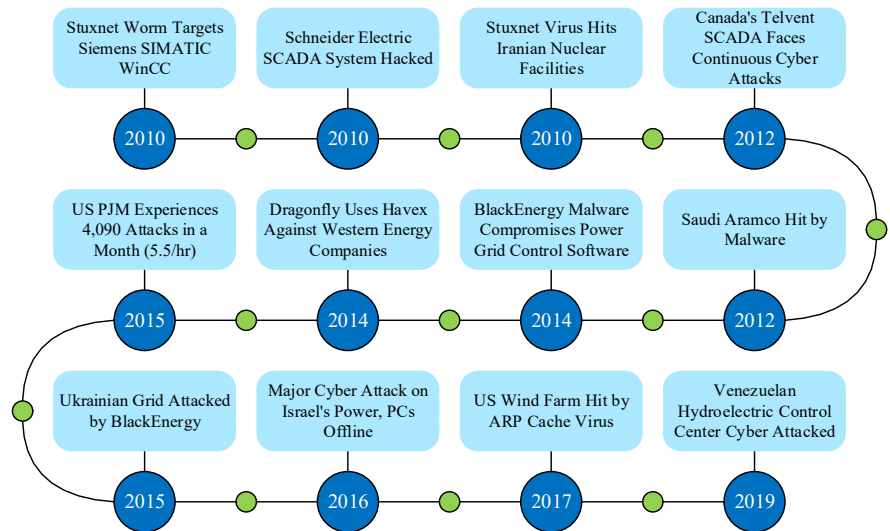

Fig. 2 Actual cases of cyber attacks against power grid

Among the various types of cyber-attacks on power CPS, False Data Injection Attacks (FDIAs) are particularly notable due to their accessibility, disruptiveness, and stealthiness [29-31]. Introduced by Liu Yao et al. in 2009 [32], FDIAs against state estimation in power grids demonstrate that attackers can infiltrate the power CPS information and communication network, gain access to network parameters and topological structures, and manipulate measurement devices to create fake data that evades bad data detection, thus misleading the control center into making erroneous operational decisions, potentially destabilizing the grid [33-37]. FDIAs in power CPS render traditional bad data detection mechanisms completely ineffective, posing a severe threat to the robust operation of the power grid [38, 39]. The typical process of a power CPS FDIA is illustrated in Figure 3.

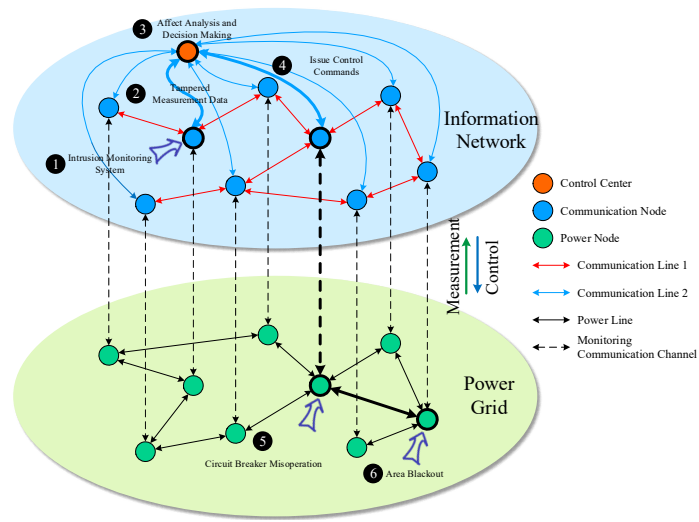

Fig. 3 Typical power CPS FDIAs process

Attackers infiltrate monitoring systems through pre-designed intrinsic attack steps, intentionally tamper with measurement data to compromise the informational integrity of the power CPS [40–42], thereby impacting the upper-level control center's analysis and decision-making, issuing control commands that lead to system switches and circuit breaker failures or misoperations, resulting in severe consequences such as widespread power outages [43-45]. Analyzing the principles and typical processes of FDIAs in power CPS allows us to summarize the characteristics of these attacks as follows:

(1) **Cross-domain Coordination:** Due to the high integration of information and physical layers in power CPS, attackers use diversified network attack methods to infiltrate the information network. By probing, elevating privileges, and controlling information systems, the attack crosses from the information domain to the power domain, ultimately impacting the physical power grid and causing physical failures in the power system across time and space [46-49].

(2) **Multi-temporal and Spatial Evolution:** FDIAs involve multiple attack steps. As time and space change, different attack entities interact at various stages, making the co-evolutionary attack process complex [50, 51].

(3) **Covert and Persistent Nature:** FDIAs, by satisfying state estimation constraints, can

successfully evade bad data detection mechanisms. Without detection, these attacks subtly influence control center decisions, leading to incorrect perceptions of the power system's operational status by control personnel. Additionally, attackers often conceal or destroy evidence after an attack to hide their tracks. Moreover, to gain extensive control, attackers typically lurk within power systems for durations ranging from several hours to days, ready to launch persistent attacks at any time [52-54].

Due to the complex characteristics and intricate evolution of FDIAs, existing attack detection methods and defense strategies struggle to effectively address FDIAs. The limitations are primarily evident in:

(1) **Lack of Characterization Methods for FDIA Evolutionary Processes:** Characterizing the spatial and temporal evolution of FDIAs could provide theoretical support for researching attack detection and data reconstruction methods. However, current achievements are mainly focused on modeling attacks such as electrical quantity manipulation, topological alteration, and GPS synchronization clock forgery, which do not suit the analysis of the temporal and spatial evolution of FDIAs [55, 56].

(2) **Challenges in Feature Extraction for Model-driven Detection Methods:** The interactive processes in power CPS are complex. Although the characteristic results analyzed by existing model-driven detection methods are reasonable and credible, the FDIA model-driven detection processes often remain in a passive detection state. Single model-driven methods struggle to comprehensively analyze and extract features, thus failing to detect FDIAs efficiently and accurately [57].

(3) **Poor Interpretability and Significant Data Influence in Data-driven Detection Methods:** Power CPS operates in a vast state space. Although existing data-driven detection methods can uncover underlying data patterns, their mechanism interpretability is poor and unconvincing. Furthermore, when system structures change, data-driven methods need time to update information. Additionally, FDIA data-driven detection methods are highly susceptible to data quality, facing severe issues such as data imbalance, high dimensionality, and difficult samples, which complicate the detection process [58].

(4) **Lack of Data Reconstruction Methods Post-FDIA Detection:** When FDIAs are detected within power CPS, the affected measurement data is often discarded, severely compromising the integrity of the measurement data. In practical power networks, directly discarding a large amount of false data may lead to unobservable local areas within the grid, creating blind spots in the network state and triggering a series of problems. Existing methods capable of reconstructing necessary data for system operation based on the remaining normal measurement data are scarce [59].

Recent years have witnessed significant advancements in the integration of cyber and physical systems within modern power grids. However, this progress has also introduced new vulnerabilities, particularly to FDIAs, which can disrupt system operations, compromise reliability, and cause cascading failures. With the increasing adoption of advanced technologies such as AI, IoT, and big data analytics, addressing these vulnerabilities has become both a technical and operational priority. Existing studies have explored various FDIA detection and defense methods, yet challenges such as the complexity of system interactions, the scalability of detection techniques, and the dynamic nature of attack scenarios remain unresolved. Moreover, the impact of FDIAs on critical aspects such as grid stability, economic efficiency, and system resilience has not been fully quantified in real-world applications. This review seeks to address these gaps by comprehensively analyzing the technological evolution, current challenges, and emerging trends in FDIA detection and defense. By establishing a power CPS coupling security analysis framework, this work aims to characterize the temporal and spatial evolution of FDIAs, develop innovative detection and reconstruction methods, and enhance the overall security and resilience of power CPS in the face of evolving cyber threats. These efforts aim to bridge the gap between theoretical advancements and practical applications, ultimately contributing to the development of more robust and intelligent power systems.

## 2 Domestic and International Research Status

This section first analyzes the data transmission scenarios and false data injection methods within power CPS, and discusses the impacts of different attack methods. It then explores the current state of research regarding the characterization of attack evolutionary processes, enhancement of attack detection training data, attack detection methods, and data reconstruction methods. Finally, it summarizes the challenges faced in research on false data injection attacks in power CPS.

### 2.1 Power CPS Data Transmission Scenarios and False Data Injection Methods

In power CPS, Phasor Measurement Units (PMUs) collect data such as nodal current phasors and voltage phasors [60], which are summarized to the Primary Domain Controller (PDC). Remote Terminal Units (RTUs), sensors, or smart meters collect real-time measurement data including nodal voltage magnitudes, reactive and active power injections at nodes, as well as reactive and active power flow on lines, which are then aggregated into data packets sent to the Supervisory Control and Data Acquisition (SCADA) system. Subsequently, the control center performs state estimation on these collected data [61, 62], which outputs unmeasurable state variables such as voltage angles and magnitudes, used for decision analysis in other applications of the Energy Management System (EMS) [63, 64]. The actual data transmission scenario and the methods of false data injection are shown in Figure 4. The injection methods can be categorized into three types:

Method 1: In-depth intrusion into the SCADA system, PDC, or communication networks to tamper with data, known as information communication network data injection attacks.

Method 2: Direct tampering with data at remote terminal devices, known as remote terminal device data injection attacks.

Method 3: Intrusion into the control center.

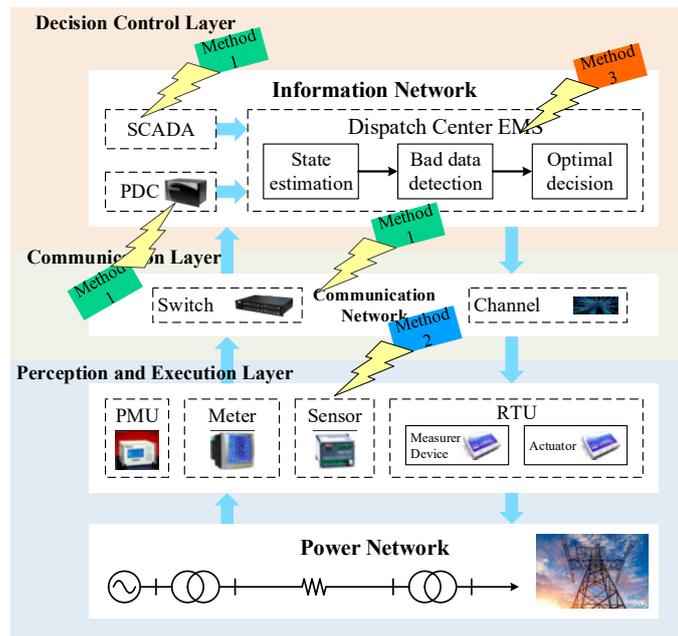

Fig. 4 Data transmission scenario and false data injection forms of power CPS

Due to the stringent security protections at actual dispatch data centers, the third method is much more difficult to execute compared to the first two methods. In power CPS, according to the principles of "security zoning, network specialization, horizontal isolation, and vertical authentication," the information communication network utilizes highly reliable, high-bandwidth, and secure fiber optic dedicated networks. The likelihood of FDIAs affecting these communication channels and devices is extremely low [65, 66]. In contrast, the lower security protection levels of public communication networks and secondary devices at grid terminals often become common points of FDIA penetration. By targeting these potential security vulnerabilities, attackers can successfully execute attacks on power CPS equipment [67-69]. Therefore, the main pathways for FDIAs are through the information communication network (method 1) and remote terminal devices (method 2).

### 2.1.1 Information Communication Network Data Injection Attacks

Data injection attacks on the information communication network represent top-level CPS attacks, typically carried out through side-channel or man-in-the-middle attacks that alter uplink measurement data or downlink control commands. This manipulation leads to control devices executing incorrect actions, thereby affecting the normal operation of the grid. The attack process typically involves three steps: 1) exploiting security vulnerabilities to infiltrate remote interfaces and control hosts to extract system control privileges; 2) using message or protocol vulnerabilities to extend control privileges to related devices; 3) remotely manipulating or disrupting the normal operation of these devices [70]. The impact of FDIAs at different stages of the information communication network is analyzed in Table 1.

Tab. 1 Impact analysis of FDIAs on different links for information and communication network

| Attack Phase | Attack Type | Attack Impact |
| --- | --- | --- |
| Software Information System | FDIAs targeting software or systems | Modification of software and hardware information |
| | FDIAs targeting control commands | Incorrect execution of control commands |
| Network Access Process | FDIAs targeting protocol vulnerabilities | Manipulation of network access data |
| | FDIAs targeting data packets | Data packet interception and tampering |
| Physical Communication Process | FDIAs targeting positioning signals | GPS positioning information spoofing |
| | FDIAs targeting time synchronization | PMU data desynchronization |

From Table 1, it is evident that such attacks can lead to alterations in software and hardware information, execution errors in control commands, manipulation of data, forgery of information, and data desynchronization, ultimately causing controlled physical devices to operate in faulty states and disrupting the normal operation of the grid.

### 2.1.2 Remote Terminal Device Data Injection Attacks

Remote terminal device data injection attacks are launched from the lower layers of the CPS. In practical power grids, devices such as smart meters, sensors, RTUs, and PMUs are not absolutely secure in network security aspects, as they are primarily designed with a focus on effective control functionality rather than network security. In certain extreme cases, terminal devices are at risk of being attacked, such as through direct physical contact [71]. Attackers exploit security vulnerabilities in these devices, breach encryption and authentication mechanisms, and inject false measurement data or control commands to achieve their objectives. The impact of FDIAs on different segments of remote terminal devices is analyzed in Table 2.

For remote terminal device FDIAs, as shown in Table 2, these attacks can cause alterations in device settings, collection errors, and external execution of control commands, impacting the control center's dispatch decisions, causing devices to execute incorrect instructions, and affecting the safety and stability of the grid's operation [72].

Tab. 2 Impact analysis of FDIAs on different links for remote terminal device

| Attack Phase | Attack Type | Attack Impact |
|---|---|---|
| Device Management Function | FDIAs targeting the device itself | Modification of device settings |
| Data Collection Process | FDIAs targeting measurement devices | Errors in collected switch and analog signals |
| Command Control Process | FDIAs targeting execution devices | Incorrect execution of control commands |

Regardless of whether it is an information communication network data injection attack or a remote terminal device data injection attack, due to different attack stages and methods, the impacts of the attacks vary. Therefore, analyzing the impacts of different FDIAs on the grid requires the development of security analysis models tailored to specific attack scenarios [73-75].

## 2.2 Current Research Status on False Data Injection Attacks in Power CPS

The detection of FDIAs and data reconstruction in power CPS start from enhancing the system's own protection capabilities [76, 77]. The goal is to interrupt attacks before they can cause severe consequences [78-80], leveraging an understanding of the attackers' objectives or behaviors to detect their actions, mitigate the actual damage to the grid, and enhance system security [81, 82]. In the research of FDIA detection and data reconstruction, four key issues need to be addressed:

### 2.2.1 Characterization of the FDIA Evolutionary Process

The essence of characterizing the FDIA evolutionary process is to analyze the temporal and spatial evolution mechanisms of FDIAs and to formally represent these processes. The goal is to develop models that are applicable for analyzing FDIA detection and data reconstruction scenarios [83-86]. Current research on FDIA evolutionary process characterization varies based on the attack scenario and includes methods for representing electrical quantity manipulation, topology alteration, and GPS synchronization clock forgery attacks, as illustrated in Tab. 3.

Tab. 3 Research status of evolutionary process characterization for FDIAs

| FDIA Evolutionary Process Characterization Methods | Attack Target | Specific descriptions |
|---|---|---|
| Electrical Quantity Manipulation Attack Characterization | Electrical quantity data collected by monitoring systems | Linear programming representation model [87] |
| | | Bilevel linear programming representation model [88] |
| | | Heuristic algorithm for solving the evolutionary representation model of an attack [89] |
| | | Sparse attack vector representation method [90] |
| | | Optimization representation method for attack-defense strategies based on a master-slave game model [91] |
| | | Feasible attack representation model constructed by minimizing angular |

|  |  | deviation of data from both sides [92] |
|  |  | Representation method for inferring system topology and parameters from cyber-physical measurement data [93] |
|  |  | Feasible attack domain representation method using a mixed integer linear programming model [94] |
| Topological Manipulation Attack Characterization | The power system network topology | Using an attack tree representation model to implement FDIA topological manipulation attacks [95] |
|  |  | Using a markov representation model to calculate the probability of attack success [96] |
|  |  | Designing attack methods that involve adding and simultaneously increasing or decreasing lines [97] |
|  |  | Fdia representation model considering power flow constraints [98] |
|  |  | Constructing attack vectors based on topology and flow data after a line break [99] |
| GPS Synchronization Clock Forgery Attack Characterization | The timestamps of PMU data | Introducing an optimal attack representation method under the constraint of positional distance differences [100] |
|  |  | Constructing an attack vector that includes the attacked PMU position and optimal phase angle manipulation values [101] |
|  |  | Developing an undetectable GPS clock attack method [102] |

(1) **Characterization Methods for Electrical Quantity Manipulation Attack Evolution:** These methods focus on the electrical quantity data collected by monitoring systems as the target of the attack. Reference [87] developed a linear programming representation model aimed at minimizing deviations and the number of changed measurements. Reference [88] introduced a dual-layer linear programming model that maximizes the consequences of an attack, calculating attack vectors under a constrained number of tampering attempts. Reference [89] utilized heuristic algorithms to solve the evolutionary representation model of the attack. Reference [90] proposed a method for representing sparse attack vectors, constrained by increasing the number of untamperable state variables. Reference [91] based on the Stackelberg game model, developed an optimized representation method for attack-defense strategies. Reference [92] constructed a feasible attack representation model that utilizes the minimization of measurement data deviation angles under unknown system parameters. Reference [93] introduced a method for inferring system topology and parameters, which requires long-term observation of data from both the information and physical sides. Reference [94] used a mixed integer linear programming model to propose a method for determining feasible attack domains with only partial system parameter information available.

(2) **Topological Alteration Attack Evolution Representation Methods:** These focus on the power system network topology as the target of the attack. Reference [95] introduced an attack tree representation model that facilitates FDIA topology alteration attacks. Reference [96] proposed a Markov representation model that enumerates FDIAs data infiltration attempts into power CPS and calculates the probability of successful attacks. Reference [97] developed a representation model targeting flow, marginal prices, and generation costs, designing attacks that increase or simultaneously increase and decrease lines, and employed a natural aggregation algorithm to solve for the attack representation model. Reference [98] aimed to increase customer electricity costs and established an FDIA representation model considering power flow constraints. Reference [99] constructed attack vectors based on topology and flow data after a line break, using false data injection to hide real line breaks, leading to more severe cascading failures.

(3) **GPS Synchronization Clock Forgery Attack Evolution Representation Methods:** These methods focus on the timestamps of synchronized phasor data collected by PMUs as the target of the attack. Reference [100] proposed an optimal attack representation method based on GPS positioning and synchronized time under the constraints of positioning distance differences. Reference [101] constructed an attack vector that includes the location of the attacked PMU and optimal phase angle tampering values, considering the principle of state estimation deviation between PMU and SCADA hybrid measurement systems. Reference [102] developed a method for undetectable GPS clock attacks, identifying one or more optimal attack targets within the PMU measurement system.

In summary, current research on FDIA evolutionary process characterization in power CPS is diverse and varied, with some focusing on attacks on electrical quantity monitoring data, others on switch quantity monitoring data and control commands, and still others on synchronized clock signal attacks. However, from any perspective, the current research focuses on attack modeling or evolutionary process characterization for specific attack targets and lacks methods that consider "data closed-loop flow characteristics" as the driving force for representing the FDIA evolutionary process. The essence of FDIA temporal and spatial evolution is the impact of the attack data stream on the information flow-energy flow conversion process in power CPS. Most existing studies model specific attack scenarios such as electrical quantity manipulation, topology alteration, and GPS synchronization clock forgery, and their representation methods are not suitable for analyzing the temporal and spatial evolution of FDIAs, nor can they provide a theoretical basis for subsequent FDIA detection and data reconstruction methods.

## 2.2.2 FDIA Detection Training Data Enhancement

Enhancing FDIA detection training data essentially involves algorithmic data balancing and redundancy reduction to improve the classification accuracy of detection models and reduce computational costs [103, 104]. Current research on FDIA detection training data enhancement varies based on the data processing method and includes oversampling, undersampling, hybrid sampling, and feature selection data enhancement methods, as shown in Tab. 4.

**Tab. 4 Research status of training data augmentation for FDIAs detection**

| Data enhancement methods | Specific descriptions | Principles |
|---|---|---|
| Over-sampling | K-nearest neighbor based SMOTE algorithm [105] | Introduce new minority samples for balance |
| | Neighborhood safety coefficient based oversampling [106] | |
| | Heilinger distance guided sample synthesis direction [107, 108] | |
| | Secondary synthetic sample strategy [109] | |
| | Adaptive synthetic oversampling algorithm [110] | |
| | Classification sorting and weight-based oversampling [111] | |
| Under-sampling | Class overlap degree-based undersampling method [112] | Remove some majority samples for balance |
| | Cluster-based undersampling method [113] | |
| | Undersampling + genetic algorithm [114] | |
| Hybrid sampling | SMOTE oversampling + EM clustering undersampling [115] | Combine oversampling and undersampling for balance |
| | SMOTE oversampling and fuzzy C-means clustering undersampling [116] | |
| | Minority oversampling + editing nearest neighbor undersampling [117] | |
| | Random undersampling + SMOTE oversampling [118] | |
| | SMOTE oversampling + clustering undersampling [119] | |
| Feature selection | Feature selection + instance selection [120] | Select relevant features for dimension reduction |
| | Firework algorithm based on feature weight selection [121] | |
| | Rough balance-based feature selection method [122] | |
| | Feature significance based feature selection method [123] | |

(1) **Oversampling Data Enhancement Methods** involve introducing new minority class samples to achieve data balance. Reference [105] addresses the issue of overgeneralization in Synthetic Minority Over-sampling Technique (SMOTE) by proposing a k-nearest neighbor-based SMOTE algorithm that assigns smaller selection weights to neighboring directions where serious overgeneralization may occur. Reference [106] introduced an oversampling method based on neighborhood safety coefficients, using inverse neighbor sampling safety coefficients to prevent newly generated data from encroaching into other classes' areas. References [107, 108] guide the synthesis of samples by comparing the Hellinger distance within the neighborhood of minority class instances and evaluate the quality of sampling. Reference [109] employs a secondary synthesis strategy, performing an initial synthesis based on support for minority class samples containing important information, followed by a second synthesis through neighborhood samples of minority class sample clusters. Reference [110] uses an adaptive synthetic oversampling algorithm for data balancing, providing different weights to minority classes to adaptively generate minority class samples. Reference [111] proposed an oversampling method based on classification ranking and weights, first sorting within-class samples based on their distance to the hyperplane, then sampling the original samples based on the density of data around the sampling points.

**(2) Undersampling Data Enhancement Methods** involve removing some majority class samples to achieve data balance. Reference [112] proposes an undersampling method based on class overlap, selecting samples that are crucial for classification based on the degree of class overlap. Reference [113] combines clustering with undersampling to propose a clustering-based undersampling method, undersampling the most informative classes by clustering majority class samples. Reference [114] introduces a genetic algorithm, combining undersampling with the genetic algorithm to achieve a balanced data processing method that trains first and balances later, obtaining multiple sets of classes with the highest information value through a single-class classifier, then using the genetic algorithm to optimize multiple random undersampled data subsets to achieve the best dataset.

**(3) Hybrid Sampling Data Enhancement Methods** combine oversampling and undersampling techniques to achieve data balance. Reference [115] mixes the SMOTE oversampling method with an Expectation Maximization (EM) clustering undersampling method, with SMOTE responsible for oversampling minority class samples and EM for undersampling majority class samples. Reference [116] combines the SMOTE method with the Fuzzy C-Means clustering method to make all classes have a similar number of instances and randomly selects instances from each cluster to achieve data balance. Reference [117] balances multiple classes based on the overlap of classes and uses minority oversampling and edited nearest neighbor methods separately for minority and majority classes. Reference [118] proposed a multiple random balancing method, using random class proportions for random undersampling and SMOTE oversampling, extending it to multiclass imbalanced datasets, using randomly generated priors for sampling. Reference [119] performs random balanced resampling of majority and minority class samples based on sample weights, using SMOTE oversampling for heavily weighted minority class samples and clustering undersampling for heavily weighted majority class samples.

**(4) Feature Selection Data Enhancement Methods** involve selecting strongly correlated features to achieve data dimension reduction. Reference [120] proposes a method combining feature selection with instance selection, using feature selection to limit features that may complicate class boundary recognition and instance selection to find the right class distribution to address imbalance and eliminate noise instances. Reference [121] introduces a fireworks algorithm for feature selection based on feature weight selection, continuously updating the optimal feature selection process through storage and selection pools. Reference [122] proposes a rough balance-based feature selection method, borrowing ideas from random subspace and random forest approaches, randomly extracting a subset of attributes from a set containing all attributes to train base classifiers. Reference [123] considers class distribution unevenness through feature significance, calculates the feature significance of each attribute based on the granular structure of each instance in the boundary region, then selects the optimal feature dataset based on feature significance.

In summary, current research on data enhancement for training detection models for false data injection attacks in power CPS is varied and competitive. However, each method has its drawbacks: 1) Oversampling methods generate minority class samples that differ from real collected samples, increasing sample diversity and quantity while introducing sample noise, which may reduce the classification accuracy for minority class samples. 2) Undersampling methods lose a large amount of majority class sample data, preventing the model from fully learning the sample features, thus reducing the accuracy of majority class sample classification. 3) Hybrid sampling methods do not show significant improvement in cases with a low imbalance ratio and have a high time complexity for training. 4) Feature selection methods have poor time performance on large datasets, and selecting features in noisy classes can reduce the generalization ability of classifiers.

### 2.2.3 FDIA Detection Approaches

FDIA detection essentially involves using data on the operational state of power CPS to determine if there are anomalies within the system and to identify whether these anomalies are caused by natural faults or by attack events [124, 125]. Current research on FDIA detection primarily utilizes state estimation, trajectory prediction, and Artificial Intelligence (AI) to detect FDIA incidents, as shown in Tab. 5.

**Tab. 5 Research status of FDIAs behavior detection**

| Detection Methods | Specific descriptions | Advantages and Disadvantages |
|---|---|---|
| State Estimation | Equivalent Measurement Transformation + Residual Detection Method [126] | Mature algorithms; fast but sensitive to threshold settings |
| | Measurement Protection Strategy + State Variable Verification [127] | |
| | Parallel Estimators + Improved State Estimation Algorithm [128] | |
| | Graph Partitioning + Chi-Square Test Method [129-131] | |
| Trajectory | Short-Term State Forecasting + Consistency Testing Method [132] | Detects false data well, but high complexity |
| | Generalized Likelihood Ratio+ High-Performance Computing [133] | |

| Prediction | Multi-Sensor Track Fusion + Particle Filtering [134] | and slow; unsuitable for complex systems |
|---|---|---|
| Artificial Intelligence | XGBoost Load Forecasting + UKF Dynamic Estimation [135] | Strong computational capabilities; clear framework; generally poor interpretability |
| | Deep Learning Techniques + Feature Extraction [136] | |
| | Batch Processing + Online Learning Algorithms [137] | |
| | Convolutional Neural Network + Model Design [138] | |
| | Equivalent Measurement Transformation+ Residual Detection [110] | |

(1) **State Estimation Detection Methods**: Reference [126] proposed a new method for FDIA detection and identification that uses equivalent measurement transformation instead of traditional weighted least squares in the state estimation process, with residual detection methods to identify FDIA. Reference [127], after analyzing the FDIA process, explored a detection method by independently verifying or measuring state variable values chosen by a set of strategic sensor measurements. Reference [128], considering the robustness of different state estimators, improved the grid state estimation attack detection by running multiple robust least squares estimators with different breakdown points in parallel, thereby enhancing the overall network security of power CPS state estimation. References [129,130] proposed a tolerable FDIA detection method based on extended distributed state estimation, using graph partitioning to divide the power grid into multiple subsystems. Each subsystem is expanded to generate extended subsystems, and the chi-square test is used to detect erroneous data in each expanded subsystem. This significantly differentiates false data from normal observational errors, thus enhancing detection sensitivity. The existing state estimation detection methods are passive, with the advantage of using mature algorithms that can reflect the characteristics of power CPS well and provide fast detection speeds [131]. However, their drawback lies in their susceptibility to detection threshold settings, which can lead to high rates of false negatives or false positives.

(2)**Trajectory Prediction Detection Methods**: Reference [132] extended the approximate direct current model to a general linear model, derived a universal FDIA model, and based on this developed a short-term state prediction method considering temporal correlations and statistical consistency testing methods to verify the consistency between predicted and received measurement values. Reference [133] proposed a generalized likelihood ratio sequence detector to address FDIA detection, which is robust against various attack strategies and load conditions in power systems, and its computational complexity is linearly proportional to the number of measurement devices, ensuring high-performance characteristics of the detector. Reference [134] introduced a multi-sensor trajectory fusion model prediction method to extract initial correlation information of attacked oscillation parameters, using a Kalman-like particle filter smoother at each monitoring node, and diagonalized this smoother into subsystems to handle continuous load fluctuations and disturbances caused by FDIA in the grid. Existing trajectory prediction detection methods are passive, with the advantage of predicting the distribution of state variables based on the historical database and operational rules of the system, matching operational trajectories, and effectively detecting various types of false data. However, their drawbacks include inapplicability to complex systems, slow detection speeds, and high computational complexity.

(3)**AI Detection Methods**: Reference [135] proposed a grid FDIA detection method based on XGBoost combined with Unscented Kalman Filter (UKF), where XGBoost load prediction results are adaptively mixed with state quantities obtained from UKF dynamic state estimation. This method uses the central limit theorem to compare the distribution of random variables for FDIA detection. Reference [138] used deep learning techniques to extract historical measurement data characteristics of FDIA behaviors and employed the captured features for real-time FDIA detection, effectively relaxing assumptions about potential attack scenarios and achieving high detection accuracy. Reference [137] combined batch and online learning algorithms (supervised and semi-supervised) with decision and feature-level fusion to build an attack detection model, analyzing the statistical and geometric properties of attack vectors used in attack scenarios and learning algorithms to detect unobservable attacks. Reference [138] constructed an FDIA detection model based on an improved convolutional neural network, implementing an efficient real-time FDIA detector based on the proposed model. Existing AI detection methods are active, with the advantage of a clear framework and strong computational power, but they suffer from poor interpretability under the complex operational mechanics of power CPS.

In summary, current research on FDIA detection in power CPS is diverse and strong in various aspects. Some focus on state estimation for detection speed, others on trajectory prediction for accuracy, and still others on AI for computational power and framework. However, each approach has its strengths and weaknesses. This paper tentatively classifies state estimation and trajectory prediction as model-driven methods, and AI approaches as data-driven methods, offering a comprehensive analysis of both: on one hand, model-driven methods are theoretically supported, producing reasonable and credible feature results. On the other hand, purely data-driven AI methods, although capable of uncovering underlying data patterns, suffer from poor interpretability and credibility, especially when system structures change and data-driven methods require time to update information. In the vast state space of power CPS, relying solely on model-driven methods makes it challenging to analyze and extract features comprehensively, whereas data-driven methods can capture features that may not yet be understood, overly simplified, or overlooked in theoretical analysis.

### 2.2.4 FDIAs Data Reconstruction in Power CPS

FDIAs data reconstruction is fundamentally a corrective protection method that uses knowledge of attack vectors or system characteristics to reconstruct damaged data and control signals, aiming to achieve desired control effects [139]. Current research on FDIAs data reconstruction is mainly divided into two categories: state-aware attack data reconstruction methods and action control attack data reconstruction methods, as illustrated in Tab. 6.

Tab. 6 Research status of FDIAs data reconstruction

| Reconstruction methods | Specific descriptions | Response Strategies |
|---|---|---|
| State Awareness Attack Data Reconstruction Method | Online GAN Measurement Data Reconstruction Method [140, 141] | Response to Attacks Targeting State Awareness |
| | Derivation of Reconstruction Matrix to Correct Attacked Angle Counters [142] | |
| | Using IGAN to Reconstruct Attacked Measurement Data [143] | |
| | Utilizing System Model to Calculate and Reconstruct Monitoring Errors [144] | |
| | Determining Mode Parameters and Reconstructing Mode Analysis Results [145] | |
| | Using SAGAN Generated Data to Restore Deceptive Data [146, 147] | |
| | Using MisGAN to Reconstruct Malicious Attack Data [148] | |
| | Using WAE Model to Restore Anomalous Data [149] | |
| Action Control Attack Data Reconstruction Method | Deriving FDIAS Signal and Its Reconstruction, Reference [150] | Response to Attacks Targeting Control Functions |
| | Adjustment Method for Feedback Controller Gain Parameters [151] | |
| | DER Attack Scenario Data Reconstruction Control Scheme [152] | |

(1) **State-Aware Attack Data Reconstruction Methods**: These methods address attacks targeting state awareness. References [140,141] proposed an online Generative Adversarial Network (GAN) measurement data reconstruction method for FDIAs, effectively reducing the impact of FDIAs on the power grid. Reference [142] introduced a method based on phase angle deviations to determine the presence of FDIAs and, based on this determination, to localize the attack and correct the affected phase angle data using a reconstruction matrix. Reference [143] utilized a Wasserstein GAN (WGAN) to reconstruct attack measurement data, achieving data integrity defense. Reference [144] employed an approximate Bayesian filter for attack vector estimation and attack detection, reconstructing monitoring errors using the system model based on dynamic power system model analysis of PMU measurement data. Reference [145] proposed a method based on spatial distribution deviation and historical bias to determine the existence of GPS spoofing attacks, deciding whether to reconstruct the pattern analysis results based on model parameters. References [146,147] improved a Self-Attention Mechanism GAN (SAGAN) through training, using the generator to restore deceptive data and enabling proactive defense against GPS spoofing attacks in smart grids. Reference [148] used a GAN trained on incomplete data (MisGAN) to reconstruct malicious attack data in a pervasive power IoT environment. Reference [149] applied a Variational Autoencoder (VAE) model to restore anomalous data to normal operating states, achieving FDIAs data reconstruction.

(2) **Action Control Attack Data Reconstruction Methods**: These methods address attacks targeting control functions. Reference [150] used an adaptive sliding mode observer to calculate errors in system control and state variables, detecting the presence of FDIAs and reconstructing the signals in case of an attack. Reference [151] proposed a method to adjust feedback controller gain parameters using an energy storage system frequency control signal as an example. Reference [152] designed a reconstruction control scheme suitable for microgrid Distributed Energy Resource (DER) attack scenarios, adopting centralized and distributed control methods for normal DER devices and those affected by attacks, respectively, and implementing additional control loops and adjusting frequency reference values to ensure stable operation and frequency control of the microgrid.

In summary, current research on data reconstruction for FDIAs in power CPS is vibrant and varied. However, there are several limitations in the existing research: 1) Electrical quantities collected by power CPS have certain characteristics of normal measurement data, but existing FDIA data reconstruction methods lack consideration for the distribution characteristics or patterns of electrical quantities; 2) Existing FDIA data reconstruction methods involve the removal, correction, or restoration of measurements, but due to computational resource limitations, convergence speed, and the dynamic variability of power systems, issues such as gradient vanishing can arise.

## 2.3 Challenges in Research on FDIAs in Power CPS

(1) **The Essence of CPS and the Impact of FDIAs**: Power CPS fundamentally involves the cyclic conversion between information flow and energy flow, while the essence of FDIA temporal and spatial evolution is the impact of the attack data stream on this conversion process. Most existing research models specific attack scenarios such as electrical quantity manipulation, topology alteration, and GPS synchronization clock forgery, with their representation methods not suitable for analyzing the temporal and spatial evolution of FDIAs. There is a lack of methods to characterize the FDIA evolutionary process from the perspective of CPS "data closed-loop flow characteristics" [153-156].

(2) **Challenges of High-Risk, Low-Frequency Events in Power CPS**: FDIAs in power CPS are "low-probability, high-risk" events characterized by imbalanced attack samples, high data dimensions, and noise. From a data-driven perspective, these characteristics are not conducive to training FDIA detection models, leading to low detection accuracy, slow real-time detection efficiency, and poor generalization capabilities [157,158].

(3) **Passive Detection Status of FDIAs and Related Issues**: Power CPS FDIAs often remain in a passive detection state. Model-driven detection methods struggle with feature extraction, and data-driven detection methods have poor interpretability. Additionally, conventional fault measurement data and FDIA measurement data have high similarity, making it difficult for single model-driven or data-driven detection methods to accurately classify challenging samples, resulting in high false positive and false negative rates [159-161].

(4) **Immediate Action Required Upon Detection**: To prevent further impact of false data on power CPS, it is crucial to promptly remove detected false data. However, this removal can severely compromise the integrity of PMU data and other measurements. In practice, when a large amount of false data is detected and removed, it can lead to unobservable local areas within the grid, thereby triggering a series of problems [162-164].

## 3 Future Research Directions

To address the challenges identified in this review and advance the resilience and security of power CPS, future research should focus on the following areas with clear technical pathways:

### 3.1 Comprehensive Characterization of FDIA Temporal-Spatial Evolution

FDIAs exhibit complex temporal and spatial evolution characteristics, significantly impacting the interaction between information and energy flows in CPS. Existing studies primarily focus on specific attack scenarios, such as electrical quantity manipulation or topology alterations. However, future research should aim to develop unified modeling frameworks that capture the broader "data closed-loop flow characteristics" of CPS. Advanced techniques, such as graph neural networks (GNNs), spatiotemporal correlation models, and causality analysis tools, can be employed to uncover latent relationships between cyber events and physical system responses. Additionally, hybrid models integrating system dynamics with real-time network behaviors could provide valuable insights into the propagation and mitigation of cascading failures caused by FDIAs.

### 3.2 Hybrid Detection Frameworks Integrating Model- and Data-Driven Approaches

Current detection methods often rely on either model-driven or data-driven approaches, each with inherent limitations. Model-driven methods provide theoretical robustness but struggle with scalability in complex systems, while data-driven techniques offer adaptability but lack interpretability. Future research should focus on hybrid frameworks that leverage the strengths of both approaches. For instance, explainable AI (XAI) can enhance the transparency of data-driven methods, allowing system operators to understand and trust detection results. Additionally, integrating domain-specific knowledge from physics-based models with machine learning algorithms can improve detection accuracy in dynamic environments. Practical implementations could include multi-layer defense architectures combining real-time anomaly detection with predictive diagnostics.

### 3.3 Advanced Data Augmentation for FDIA Detection

Training robust detection models requires addressing the imbalance and high-dimensional nature of FDIA datasets. Existing methods, such as oversampling and undersampling, often fail to preserve the statistical characteristics of minority-class samples. Future research should explore advanced techniques, such as generative adversarial networks (GANs), to create realistic synthetic data for training. Federated learning frameworks could enable multiple entities to collaboratively train detection models without compromising data privacy. Furthermore, automated feature engineering methods, including dimensionality reduction and clustering algorithms, can help extract meaningful patterns from high-dimensional datasets, improving detection performance.

### 3.4 Resilient Data Reconstruction Techniques

When FDIAs compromise measurement data, ensuring the integrity and availability of system information is critical. Current reconstruction methods often ignore the dynamic variability of CPS data and may lead to system observability gaps. Future research should develop probabilistic models, such as VAEs, and self-supervised learning techniques to recover missing or corrupted data while preserving its statistical and temporal characteristics. Incorporating robust optimization methods, such as reinforcement learning, can enhance real-time decision-making during reconstruction. Case studies on practical power grid scenarios will help validate these methods and improve their applicability.

### 3.5 Information Security in Integrated Energy Systems

Integrated energy systems (IESs) represent the convergence of electricity, gas, heat, and renewable energy resources [165-168], creating unique cybersecurity challenges. Unlike traditional power grids, an IES involves uncertain renewable energy resources [169, 170], together with multiple domains with varying communication protocols and security vulnerabilities. Future research should develop cross-domain security frameworks that address these challenges by leveraging blockchain for secure transactions, digital twins for real-time threat simulations, and multi-layered defense strategies to mitigate cascading attacks. For instance, adaptive intrusion detection systems could analyze communication patterns across domains, enabling early detection of cyberattacks. Collaborative efforts between researchers and industry stakeholders are essential to develop scalable solutions tailored to the complexity of IESs.

### 3.6 Integration of Emerging Technologies

Emerging technologies offer transformative opportunities to enhance CPS and IES security. Blockchain can ensure tamper-proof records for critical data exchanges [171], while digital twins can model and simulate system behaviors to anticipate vulnerabilities [172]. Data Encryption is able to enchane the security of CPS [173]. In addition, advanced AI is capable of optimizing the secure and economic operation of CPS and identifying appliances' behaviors through appliance-specific networks [174, 175]. Quantum computing, with its unparalled computational capabilities, could revolutionize attack detection and optimization methods by accelerating the resolution of complex problems [176]. Future research should focus on integrating these technologies into practical systems, emphasizing interoperability, scalability, and cost-effectiveness. Pilot projects demonstrating their real-world applications in smart grids and multi-energy systems will be instrumental in gaining broader acceptance.

**3.7 Policy and Standardization**

The development of robust technical solutions must be complemented by clear regulatory frameworks and international standards to ensure widespread adoption. Future research should explore the interplay between technology and policy, focusing on areas such as data privacy, cross-border energy trading, and incident response protocols. For example, establishing guidelines for cybersecurity audits and compliance in CPS and IES environments can promote trust among stakeholders. Collaborative efforts between academia, industry, and regulatory bodies are essential to develop cohesive strategies that balance innovation with security requirements.

# 4 Conlcusion

This review has comprehensively examined the detection, evolution, and data reconstruction strategies for FDIAs in CPS. By analyzing the temporal and spatial evolution of FDIAs, this work has highlighted the significant vulnerabilities introduced by the integration of cyber and physical domains in modern power systems. A critical assessment of existing detection methods has revealed gaps in addressing the scalability, interpretability, and resilience of current solutions. Similarly, data reconstruction approaches were evaluated, underscoring the challenges of maintaining data integrity and system observability during and after an attack.

Future research must focus on developing unified frameworks that incorporate advanced modeling techniques, such as spatiotemporal analysis and machine learning, to better characterize FDIA evolution and enhance detection accuracy. The integration of emerging technologies, including blockchain, digital twins, and quantum computing, also holds promise for improving system resilience and operational security.

By addressing the identified challenges, this work aims to bridge theoretical advancements with practical applications, contributing to the development of robust, intelligent, and secure power CPS capable of withstanding evolving cyber threats. These efforts not only safeguard the reliability of power grids but also provide a foundation for the broader adoption of integrated energy systems.